\documentstyle[multicol,aps,prl,epsfig]{revtex}
\begin{document}
\topmargin=0.01cm
\title{Manifestation of anisotropy persistence in
the hierarchies of MHD scaling exponents}
\author{N.\,V.~Antonov,$^{1}$ J.~Honkonen,$^{2}$
A.~Mazzino$^{3}$ and P.~Muratore-Ginanneschi$^{4}$\\
\small{$^{1}$
Department of Theoretical Physics, St Petersburg University,
Uljanovskaja 1, St Petersburg, Petrodvorez, 198904 Russia.\\
\small$^2$ Theory Division, Department of Physics,
P.O.Box 9, FIN-00014 University of Helsinki, Finland.\\
\small$^3$ INFM--Department of Physics, University of Genova, I--16146
Genova, Italy.}\\
\small$^4$ The Niels Bohr Institute, Blegdamsvej 17, DK-2100 Copenhagen,
Denmark}
\draft
\date{\today}
\maketitle
\begin{abstract}
The first example of a turbulent system where the failure of the
hypothesis of small-scale isotropy restoration
is detectable both in the `flattening' of
the inertial-range scaling exponent hierarchy,
and in  the behavior of odd-order dimensionless ratios, e.g., skewness and
hyperskewness, is presented.
Specifically, within the kinematic approximation in magnetohydrodynamical
turbulence, we show that for compressible flows,
the isotropic contribution to the scaling of magnetic correlation functions
and the first anisotropic ones may become practically indistinguishable.
Moreover, skewness factor now diverges as the P\'eclet number goes to
infinity, a further indication of small-scale anisotropy.
\end{abstract}
\pacs{PACS number(s)\,: 47.27.Te, 47.27.$-$i, 05.10.Cc}

\begin{multicols}{2}

A wide interest has recently been devoted to the
possible occurrence of small-scale isotropy restoration
for scalar (see e.g., Ref.~\cite{SS99} and references therein),
Navier-Stokes \cite{BoOr96,Arad99} and
magnetohydrodynamical (MHD) turbulence \cite{LM99,ABP99,ALM99}.
The scenario can be summarized as follows.
In the presence of  anisotropic large-scale injection mechanisms,
the inertial-range statistics is characterized by an infinite hierarchy
of scaling exponents; however the leading contribution to scaling
comes from the isotropic component. From this point of view, one might
argue that large-scale
anisotropy does not affect inertial-range scaling properties. Actually,
focussing on a larger set of observables, small-scale anisotropies
become manifest. It turns out that the behavior of
odd-order dimensionless
ratios (e.g., skewness and hyperskewness) is completely different
from the case of small-scale isotropy restoration.
Such indicators go to zero down to the inertial range much slower than
predicted by dimensional considerations \cite{P96} or, more dramatically,
they diverge at the smallest scales \cite{ALM99} (see also \cite{A98,A99}).

The main aim of this Letter is to present a model
of MHD turbulence where, by varying the degree of
compressibility of the velocity field, anisotropic persistence
is now detectable both from the
`flattening' of the hierarchy of inertial-range scaling
exponents (the isotropic component and the first anisotropic ones may
become practically indistinguishable)
and the divergence of skewness factor with the P\'eclet  number.
We give the basic ideas and results;
longer and more exhaustive technical discussions
will be  presented in a forthcoming paper.

In the presence of a mean component $\bbox{B}^o$ (actually
varying on a very large scale $L$)
and for the compressible velocity field $\bbox{v}$,
the kinematic MHD equations describing the evolution of the fluctuating
(divergence-free) part $\bbox{B}$ of the magnetic field
are \cite{Zeldo83}:
\begin{eqnarray}
\label{fp}
&&\partial_t B_\alpha +\bbox{v}\cdot\bbox{\partial}\,B_\alpha
=-(B_{\alpha}+B^o_{\alpha})\bbox{\partial}\cdot\bbox{v}+
\bbox{B}\cdot\bbox{\partial}\,v_{\alpha}+\nonumber\\
&&\qquad \bbox{B}^o\cdot \bbox{\partial}\,v_{\alpha}+
\kappa_{0}\,\partial^2 B_{\alpha},\qquad
\alpha=1,\cdots ,d ,
\end{eqnarray}
where $\kappa_0$ is the magnetic diffusivity. The field $\bbox{B}^o$
plays the same role as an external forcing driving the system and
being also a source of anisotropy for the magnetic field statistics.

Our choice for the velocity statistics generalizes that of the well-known
kinematic Kazantsev--Kraichnan model for the compressible case: $\bbox{v}$
is a Gaussian process of zero average, homogeneous, isotropic
and white in time. It is self-similar and defined by the two-point
correlation function:
\begin{equation}
\langle  v_{\alpha}(t,\bbox{x}) v_{\beta}(t',\bbox{x}') \rangle =
\delta (t-t')\, [ d^0_{\alpha\beta}
- S_{\alpha\beta} (\bbox{x}- \bbox{x}') ],
\label{2-point-v}
\end{equation}
where $d^0_{\alpha\beta} = {\rm const} \,\, \delta _{\alpha\beta}$
and $S_{\alpha\beta}(\bbox{x}-\bbox{x}')$
is fixed by isotropy and scaling:
\begin{equation}
S_{\alpha\beta}(\bbox{r})=r^{\xi}\left [{\cal X}\delta_{\alpha\beta} +
{\cal Y}\frac{r_{\alpha}r_{\beta}}{r^2} \right ]\, , \qquad r\equiv
\left |\bbox{x}-\bbox{x}' \right |\, ,
\label{eddydiff}
\end{equation}
with the coefficients
\begin{equation}
{\cal X}=\frac{{\cal S}^2(d+\xi-1)-\xi {\cal C}^2}{(d+\xi)(d-1)\xi}
\, ,\qquad
{\cal Y}=\frac{d{\cal C}^2-{\cal S}^2}{(d+\xi)(d-1)}.
\label{xy}
\end{equation}
The degree of compressibility is thus controlled by the ratio
$\wp\equiv {\cal C}^2/{\cal S}^2$, with ${\cal S}^2
\propto \langle (\partial_{i} v_{k} \partial_{i} v_{k}) \rangle $
and ${\cal C}^2\propto\langle(\bbox{\partial}\cdot\bbox{v})^2\rangle$.
It satisfies the inequality $0\leq \wp \leq 1$;
$\wp=0$ and 1 corresponding to the purely solenoidal and potential
velocity fields, respectively.

In the present Letter our attention will be focused on the
inertial-range behavior of magnetic correlation functions, where
power laws are expected in their decompositions on a set of orthonormal
functions $P$
\begin{equation}
\langle B_{\parallel}^n (t, \bbox{x}) B_{\parallel}^q (t,\bbox{x}') \rangle
={\textstyle\sum_{j=0}^{\infty}} P_j(\cos\phi)\, r^{\zeta_j^{n,q}},
\label{scaling}
\end{equation}
$B_{\parallel}$ being some component of $\bbox{B}$, e.g., its
projection along the direction
$\hat{\bbox{r}}\equiv\bbox{r}/r$ or
$\hat{\bbox{B}}^o\equiv\bbox{B}^o/B^{o}$,
and $\phi$ is the angle between $\bbox{r}$ and $\bbox{B}^o$.
Notice that, due to the anisotropic injection mechanism, the inertial-range
statistics is now characterized by an infinite hierarchy of
exponents ($j$ denotes the $j$-th anisotropic sector), rather than just one
exponent as in the isotropic case.

If the Kolmogorov 1941 isotropization hypothesis holds,
the contribution from the anisotropic sectors (i.e.~$j\neq 0$ )
to the scaling of correlation functions
should be negligible with respect to the isotropic component.
Such picture indeed holds for even correlation functions and
solenoidal velocity  \cite{LM99,ALM99},
but it breaks down for  compressibility strong enough.\\
In order to prove this fact, 
consider first the pair correlation function,
$C_{\alpha\beta}(t,\bbox{r})
\equiv\langle B_{\alpha} (t,\bbox{x}) B_{\beta}(t,\bbox{x}')\rangle$.
Here, exploiting the zero-mode technique
\cite{GK95,CFKL95,SS96,V96}, the complete set of scaling exponents
$\zeta_j^{1,1}$ can be found nonperturbatively. We 
give the basic ideas of the strategy.

The first key point is that exact, closed equation for
$C_{\alpha\beta}$ can be found due to the
time decorrelation of the velocity field:
\begin{eqnarray}
 &&\partial_t\, C_{\alpha\beta}=
S_{ij}\,\partial_i\partial_j \,C_{\alpha\beta}-
 \left (\partial_j\,S_{i\beta}\right )
 \partial_i\,C_{\alpha j}-\nonumber\\
&&\qquad\left (\partial_jS_{\alpha i}\right )\left
 (\partial_i C_{j \beta}\right )+
 \left (\partial_i\partial_j S_{\alpha\beta}\right )
 \left (  C_{ij} + B_i^oB_j^o\right ) + \nonumber \\
 &&\qquad 2\kappa_0\, \partial^2 C_{\alpha\beta}-B_\beta^oB_j^o
\partial_i\partial_j
 S_{\alpha i}-B_\alpha^oB_j^o\partial_i\partial_j S_{\beta
 i}+\nonumber\\
&&\qquad B_\alpha^oB_\beta^o\partial_i\partial_j S_{i j}-
C_{\alpha j}\,(\partial_i\partial_j \,S_{\beta i})+C_{\beta j}\,
\partial_i\partial_j \,S_{\alpha \beta}+\nonumber\\
&&\qquad C_{\alpha \beta
 }\,\partial_i\partial_j\,S_{ i j}+2(\partial_i C_{\alpha \beta})\,
\partial_j \,S_{i j}
\label{eq-moto}
\end{eqnarray}
(see \cite{ALM99} for the derivation in the incompressible case).\\
$C_{\alpha\beta}$ can be projected on a vector basis including the large scale
magnetic field $\bbox{B}^o$ \cite{LM99,ALM99}, or, alternatively, 
on the purely inertial range basis that span the irreducible representations
of SO($d$) \cite{ALP99}. Deep in the inertial range,
both descriptions yield a system 
of linear algebraic equations for the scaling law coefficients. 
The leading solutions are associated with the homogeneous 
solutions of such system, i.e.~with zero modes. Their scaling exponents
follow from the imposition that the determinant of the coefficients be zero.
Schematically, the solutions can be expressed in the form:
\begin{equation}
\zeta_j^{1,1}=\alpha+\sqrt{\beta+\gamma\sqrt{\delta}}\qquad (j\
\mbox{even}),
\label{zero-modes}
\end{equation}
where $\alpha$, $\beta$, $\gamma$ and $\delta$ are cumbersome
functions of $\xi$, $d$ and $j$ and will not be reported here for
the sake of brevity. Contributions to the scaling associated with odd $j$'s
vanish due to the symmetry $\phi\mapsto -\phi$.\\
In particular, the expression derived in
Ref.~\cite{RK97} in the isotropic case has been recovered here for $j=0$.\\
 For $j=0$ and $j=2$, the limit $\xi\to 0$ in Eq.~(\ref{zero-modes}) yields:
\begin{eqnarray}
&&\zeta_0^{1,1} = (-1+2\ \wp -d\ \wp)\ \xi +O(\xi^2),
\label{small-xi1}\\
&&\zeta_2^{1,1} = \frac{2-\wp\ [4+d\ (d-2)\ (d+1)]}{(d-1)(d+2)}\xi
+O(\xi^2),
\label{small-xi2}
\end{eqnarray}
while for large $d$ we have:
\begin{eqnarray}
&&\zeta_0^{1,1} =-\frac{\wp\ \xi}{1+\wp\ \xi}d
 -\frac{1-2 \wp\ (1-\xi)}{1+\wp\ \xi}\xi +O(1/d),\label{large-d1}\\
&&\zeta_2^{1,1} =
-\frac{\wp\ \xi}{1+\wp\ \xi}d -\frac{\wp\ \xi(\xi-2)}{1+\wp\ \xi}
+O(1/d) .
\label{large-d2}
\end{eqnarray}

\begin{figure}
\centerline{\psfig{file=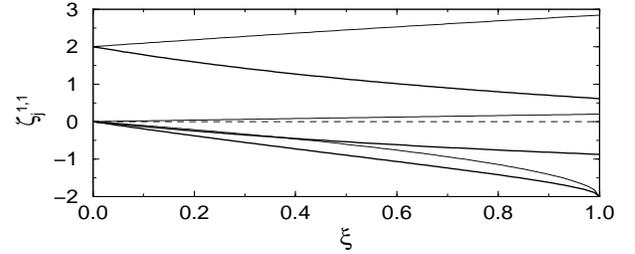,height=3.5cm,width=8cm}}
\caption{
Behavior of $\zeta_j^{1,1}$
{\it vs} $\xi$ for $d=3$ and  $j=0$, $2$ and
$4$ \mbox{\hspace{9.5cm}}
(from below to above).
Heavy lines: $\wp = 1$; thin lines: $\wp = 0$.}
\label{fig1}
\end{figure}

In Fig.~\ref{fig1} we present the behavior of $\zeta_j^{1,1}$
obtained from the expression (\ref{zero-modes})
for $\wp=0$ (thin lines) and $\wp=1$ (heavy lines). As one can see,
for $\wp=0$ we have $\zeta_0^{1,1}<0$ but $\zeta_2^{1,1}>0$ and
$\zeta_4^{1,1}>0$ so that, in particular,
$(r/L)^{\zeta_0^{1,1}}\gg (r/L)^{\zeta_2^{1,1}}$ 
in the inertial range of scales.
%
%
%
%
%
%
The situation changes for $\wp>\wp_{c}$ ($\wp_{c}\sim 0.1274$
for $j=2$ and $d=3$), when $\zeta_2^{1,1}$ becomes negative
for all $0\le\xi\le 1$. This means that, for compressibility strong enough,
$(r/L)^{\zeta_0^{1,1}}$ and $(r/L)^{\zeta_2^{1,1}}$ become very close
(still, $\zeta^{1,1}_{j}  <  \zeta^{1,1}_{k} $  for $j<k$).
As we can see from Eqs.~(\ref{large-d1}), (\ref{large-d2}) the effect
becomes
dramatic for large dimensions and $\wp \neq 0$ when
$\zeta_0^{1,1}\sim \zeta_2^{1,1}$. The contribution
to the scaling of the correlation functions in Eq.~(\ref{scaling}) coming
from the sector $j=2$ thus becomes less and less subleading
as $\wp$ and/or $d$ increase.

Further strong evidence of the crucial role of compressibility in the
failure
of the small-scale isotropy restoration
can be obtained by looking at the higher-order
correlation functions [i.e.~$n$ and/or $q>1$ in Eq.~(\ref{scaling})]
and, in particular, at dimensionless ratios of odd-order moments.
Were isotropy restored at
small scales, such ratios would go to zero
for large P\'eclet  number. The latter is defined as
Pe $\equiv(L/\eta)^{1/\xi}$, where $L$ is the integral scale and
$\eta\propto\kappa_0^{1/\xi}$ is the dissipation scale.

Non-zero values of such indicators are thus the signature of anisotropy
persistence. As we are going to show, in contrast to the
incompressible case, for $\wp$ large enough the skewness factor now
diverges
as $\mbox{Pe}\to\infty$. To be more specific, the leading contribution
in the expression (\ref{scaling}) can be written as:
\begin{equation}
\langle B_{\parallel}^n (t, \bbox{x}) B_{\parallel}^q (t,\bbox{x}') \rangle
\propto (r/\eta)^{\alpha_0^{n,q}} \, (r/L)^{\beta_0^{n,q}}
\propto r^{\zeta_0^{n,q}}\, ,
\label{final}
\end{equation}
where we have defined
$\zeta_0^{n,q}\equiv \alpha_0^{n,q}+\beta_0^{n,q}$.
The expressions for $\zeta_0^{n,q}$ have been
obtained up to first order in $\xi$  by means of the field 
theoretic renormalization group
and operator product expansion. A detailed presentation of these
techniques for the case $\wp=0$ can be found in \cite{ALM99}
(see also Refs.~\cite{A98,A99,AAV98} for the scalar case and
\cite{UFN} for general review);
below we confine ourselves to only the necessary information.

The key role is played by the scaling dimensions $\Delta[n,j]$
of the $j$-th rank composite operators
$ B_{\alpha_{1}}(x)\cdots B_{\alpha_{j}}(x)
\bigl[B_{\alpha}(x)B_{\alpha}(x)\bigr]^{l}$ ($n\equiv 2l+j$
is the total number of $\bbox{B}$'s).
The first order in $\xi$ (one-loop approximation) yields
\end{multicols}
\begin{equation}
\Delta[n,j]=\frac{\xi}{2(d-1)(d+2)} \, \left\{
n(n-1) \left[2-\wp(d^{3}-d^{2}-2d+4)\right] - (n-j)(n+j+d-2)(d+1-2\wp)
\right\} +O(\xi^{2}).
\label{Dnp}
\end{equation}
\begin{multicols}{2}
The exponents in Eq. (\ref{scaling}) are related to the dimensions
(\ref{Dnp}) as follows (see Ref. \cite{ALM99}):
$$\zeta^{n,q}_{j} = \Delta[n+q,j] - \Delta[n,j_{n}] - \Delta[q,j_{q}], $$
where $j_{n} = 0 $
for $n$ even and 1 for $n$ odd. We thus conclude that the inequality
$\partial \Delta[n,j] / \partial j \ge 0$,
which follows from Eq. (\ref{Dnp}) for all $\wp$ and $d\ge2$,
generalizes the hierarchy discussed above to the higher-order functions.
%
It becomes flatter and flatter as $\wp$ grows
($\partial^{2} \Delta[n,j] / \partial j \partial\wp \le 0$), while
for $d\to\infty$ (and $\wp \neq 0$) the effect becomes even stronger: 
in the leading
$O(d)$, expressions (\ref{Dnp}) and (\ref{zeta}) are now independent of
$j$.

The leading term of Eq. (\ref{scaling}) in the inertial range is given
by the contribution with the minimal $j$:
\begin{equation}
\zeta_0^{n,q} = \cases{
-\xi{\displaystyle
\frac{nq (1+\wp d^{2}-2\wp)+ (d+1-2\wp)}{(d+2)}} &  \cr
{\displaystyle -\frac{\xi nq }{(d+2)}}
(1+\wp d^{2}-2\wp), & \cr}
\label{zeta}
\end{equation}
where the first holds if both $n$ and $q$ are odd, and the second
otherwise.
%
%
For $n=q=1$, expression (\ref{small-xi1}) is recovered.
Knowing the exponents $\alpha_0^{n,q}$ and $\beta_0^{n,q}$,
dimensionless ratios of the form
$R_{2n+1}(r) \equiv \langle B_{||}^{2n} (\bbox{x}) B_{||} (\bbox{x}')
\rangle/\langle B_{||} (\bbox{x}) B_{||} (\bbox{x}')\rangle ^{(2n+1)/2}$
can be constructed and, as in Ref.~\cite{P96}, evaluated
at the dissipative scale [i.e.~$r=\eta$ in Eq. (\ref{final})].
When doing this, the explicit dependence on Pe
appears and the final $O(\xi)$ expressions read:
\begin{equation}
R_{2n+1}(\mbox{Pe}) \propto \mbox{Pe}^{\sigma_{2n+1}}, \qquad n=1,2,\cdots,
\label{skew}
\end{equation}
with the exponents
\begin{equation}
\sigma_{2n+1}(\wp )\equiv \frac{2 n^{2} (1-2\wp+\wp d^{2})}{(d+2)} -\frac{
(1-2\wp +\wp d)}{2}.
\label{expon}
\end{equation}
It is easy to verify from expression (\ref{expon}) that for $d\geq 2$
and $n\geq 1$
we have $\partial\sigma_{2n+1}(\wp)/\partial \wp>0$. Negative values of
$\sigma_{2n+1}(0)$ may thus become positive due to $\wp$. In particular,
for $d=3$ we obtain $\sigma_{3}(\wp)=(23\wp -1)/10$,
which becomes positive for $\wp > 1/23$. It then follows that
$R_{3}\to\infty$ as $\mbox{Pe}\to\infty$, the footprint
of the persistence of the small-scale anisotropy.

Some remarks on the limit of large space dimensions are worth.
One immediately realizes from Eq.~(\ref{zeta}) that,
for $d\to\infty$ the scaling exponents reduce to $\zeta_0^{n,q}=-\xi nq d\wp$,
that means $\zeta_0^{n,q}=n\zeta_0^{1,q}=q\zeta_0^{n,1}$,
and thus the vanishing
of intermittency for $d\to\infty$. Nevertheless, in this limit
$\zeta_0^{1,1}=\zeta_2^{1,1}$ and, at the dissipative scale $\eta$,
$\sigma_{2n+1}(\wp)=\wp d(2n^2-1/2)>0$,
two clear signatures of small-scale anisotropy persistence.
The latter may thus occur also in the absence of intermittency.

Let us now examine the possible mechanisms at the origin
of the small-scale anisotropy persistence
in our problem. They
can be easily grasped in two dimensions; our previous results being
valid for all $d\geq2$, it is reasonable to expect that 
the mechanisms we are going to show hold also for higher $d$'s.
Let us start from the 
incompressible case assuming, without loss of generality, that
$\hat{\bbox{B}}^o$
is oriented along the $y$-axis (i.e.~$\hat{\bbox{B}}^o \equiv \bbox{e}_y$).
As we shall see, all our considerations
will hold,  {\it a fortiori}, in the compressible case.

It can be shown that
the magnetic field can be represented by
the (scalar) magnetic flux function in the form
$\bbox{B}=\bbox{\partial}\psi \times\bbox{e}_z$,
where $\psi$ satisfies the
passive scalar equation forced by a large scale gradient
$\bbox{G}\equiv \bbox{B}^o\times \bbox{e}_z$. One finds
\begin{equation}
\partial_t \Psi +\bbox{v}\cdot\bbox{\partial}\,\Psi =
 \kappa_{0}\,\partial^2 \Psi
\label{duedim}
\end{equation}
where $\Psi(t,\bbox{x})\equiv \psi+ \bbox{G}\cdot \bbox{x} $.
Notice that, fort $d=2$, expression (\ref{zero-modes}) recovers the 
results of Ref.~\cite{CFKL95} obtained directly for the scalar problem.

An interesting
feature recently recognized for the passive scalar problem, for
both synthetic advection by
$\delta$-correlated velocity \cite{FGLP97} and Navier--Stokes advection
\cite{P94,HS94,CLMV99,CLMV99bis}, is related to the formation of
``cliffs'', i.e.~very steep scalar
gradients within very short distances.
The region where such gradients develop are
separated by ``plateaus'' where the scalar
 depends smoothly on the position (i.e.~regions of gradient expulsion).\\
The emergence of this ubiquitous
pattern is explained by considering 
the action of the velocity derivative matrix 
$\partial_{\alpha}v_{\beta}$.
As emphasized in Refs.~\cite{P94,HS94},
scalar gradients are weak in the elliptic regions of the velocity field
(in two dimensions they correspond to purely imaginary
eigenvalues of the velocity derivative
matrix) where its `rotational' character
inhibits the formation of strong scalar gradients. On the
contrary, in the hyperbolic regions the flow is almost `irrotational':
the alignment of scalar gradients
with the direction of the eigenvector corresponding to the most negative
eigenvalue of the velocity
derivative matrix (roughly, the direction corresponding
to compression along one direction) is not discouraged, and actually
observed \cite{P94}, and strong scalar gradients develop.

When the scalar is forced by isotropic injection mechanisms, no
preferential direction arises and the intense gradients are randomly
oriented. The final result is a small scale isotropic statistics.
The situation changes in the case of non-isotropic injection mechanisms
like the one encountered here. In this case strong gradients
are oriented  along $\bbox{G}$ and,
as very recently shown in  Refs.~\cite{CLMV99,CLMV99bis},
small-scale isotropy is consequently not restored for the scalar field.
Exploiting the relation between
the magnetic field and the magnetic flux function, we can
conclude that extreme magnetic fluctuations have a tendency to occur
preferentially along the direction $\hat{\bbox{B}}^o$,
the origin  of the observed small-scale anisotropy.

In the compressible case both eigenvalues can be negative
Compression may thus occur
in both directions, enhancing
the formation of fronts in the magnetic flux function.

In conclusion, we presented a simple model
of MHD turbulence where, by varying the degree of
compressibility of the velocity field, the persistence of anisotropy
is detectable both from the hierarchy of inertial range
exponents and from the divergence of skewness factor with the P\'eclet
number. Although our results were obtained on the base of a specific model,
they seem to be rather general: a similar behavior is also observed for a
passive scalar advected by a compressible velocity field \cite{A99}
(however, the flattening is less pronounced for the latter).

The results presented here give
the first evidence of a turbulent system displaying such
twofold behavior.

We are deeply grateful to L.\,Ts.~Adzhemyan, L.~Biferale, A.~Celani,
A.~Lanotte, and M.~Vergassola for discussions throughout the present work.
N.\,V.\,A. acknowledges the hospitality of the University of Helsinki.
N.\,V.\,A. was supported by the Academy of Finland,
GRACENAS Grant No. 97-0-14.1-30, and RFFI Grant No.~99-02-16783.
A.\,M. was partially supported by the INFM PA project No.~GEPAIGG01.
P.\,M.\,G. was partially supported by grant ERB4001GT962476 from EU.

\end{multicols}
\end{document}